\documentstyle[12pt]{article}
\input epsf.tex
\input fps.sty
\newcommand{\pat}{\partial}
\newcommand{\be}{\begin{equation}}
\newcommand{\ee}{\end{equation}}
\newcommand{\bea}{\begin{eqnarray}}
\newcommand{\eea}{\end{eqnarray}}

\newcommand{\Lcal}{{\cal L}}

\newcommand{\half}{\frac{1}{2}}

\newcommand{\Tr}{{\rm Tr}}
\newcommand{\tr}{{\rm tr}}

\newcommand{\unitm}{{\bf 1}}
\newcommand{\zerom}{{\bf 0}}

\newcommand{\bra}{\langle}
\newcommand{\ket}{\rangle}

\newcommand{\mapb}{\longleftarrow \leftharpoondown \rightharpoondown 
                   \longrightarrow}
\newcommand{\mapc}{\longleftarrow \leftharpoondown \leftharpoondown
                     \leftharpoondown \leftharpoondown
                   \rightharpoondown \rightharpoondown
                  \rightharpoondown \rightharpoondown \longrightarrow}

\begin{document}

\baselineskip 16pt

\begin{titlepage}
\begin{flushright}
CALT-68-2169 \\
hep-th/9804067 \\
April 1998
\end{flushright}

\vskip 1.2truecm

\begin{center}
{\Large {\bf M-momentum Transfer between Gravitons,}}
\end{center}
\vspace{-0.3cm}
\begin{center} 
{\Large {\bf Membranes, and Fivebranes}}
\end{center}
\vspace{-0.3cm}
\begin{center}
{\Large {\bf  as Perturbative Gauge Theory Processes}} 
\end{center}

\vskip 0.8cm

\begin{center}
{\bf Esko Keski-Vakkuri}$^1$ and {\bf Per Kraus}$^2$
\vskip 0.3cm
{\it California Institute of Technology \\
     Pasadena CA 91125, USA  \\
     email: esko or perkraus@theory.caltech.edu}
\end{center}

\vskip 1.6cm

\begin{center}
{\small {\bf Abstract: }}
\end{center}
\noindent
{\small 
Polchinski and Pouliot have shown that M-momentum transfer  between membranes
in supergravity
can be understood as a non-perturbative instanton effect in gauge theory. 
Here we consider a dual process: electric flux transmission between D-branes.
We show that this process can be described in perturbation theory as
virtual string pair creation, and is closely related to Schwinger's
treatment of the pair
creation of charged particles in a uniform electric field. 
Through the application of dualities, our perturbative calculation gives 
results for various  non-perturbative amplitudes, including M-momentum 
transfer between gravitons, membranes and longitudinal fivebranes.
Thus perturbation theory plus dualities are sufficient to demonstrate
agreement between supergravity and gauge theory for a number of M-momentum 
transferring processes. A variety of other
processes where branes are transmitted between branes, {\em e.g.}
$(p,q)$-string transmission in IIB theory, can also be studied.
 We  discuss the implications of our results for proving the  11 dimensional 
Lorentz invariance of Matrix theory.
}
\rm
\vskip 2.2cm

\small
\begin{flushleft}
$^1$ Work supported in part by a DOE grant DE-FG03-92-ER40701.\\    
$^2$ Work supported in part by a DOE grant DE-FG03-92-ER40701 and by the
DuBridge Foundation.
\end{flushleft}
\normalsize 
\end{titlepage}

\newpage
\baselineskip 18pt

\section{Introduction}

In the flat, eleven dimensional limit of M-theory, scattering amplitudes
involving the exchange of momentum in the eleventh direction are simply
related by rotational invariance to processes involving purely transverse
momentum exchange.  However, in the gauge theory description of 
M-theory used in Matrix Theory \cite{BFSS},
this rotational symmetry is made highly non-manifest and 
 must arise from relatively complicated non-perturbative
effects, if it is there at all.  Thus while numerous checks have been made
of the correspondence between transverse momentum exchange and perturbative
gauge theory processes,  so far the successful description of M-momentum
transfer has been limited to a single case involving the scattering of
membranes \cite{PP,DKM}.  In this example, it was shown by 
Polchinski and Pouliot 
that including instanton effects
in the relevant $2+1$ super Yang-Mills theory\footnote{For additional
discussion of the action used in \cite{PP}, see \cite{us} 
and De Boer {\em et. al.} in \cite{DKM}.} rendered the eleventh 
dimension on an equal footing with the other ten.  A T-dual version
of this process was considered by Banks {\em et. al.}  \cite{BFSS2}, who
 showed that by wrapping the membranes on a torus which shrinks 
to zero size one obtains Y-momentum transfer between gravitons in IIB theory.
Unfortunately, it is a
difficult and unsolved problem to carry out the analogous calculations for
cases other than membranes.

   On the other hand, since the relation between
the transverse and eleventh dimensions in M-theory maps on to the
relation between perturbative and non-perturbative effects in gauge theory,
we might hope to make further progress by application of duality 
transformations.  Many non-perturbative results in the gauge theory can
be obtained by starting from a perturbative result and applying  duality.
 In the present work we exploit this approach in order to
1) rederive the result of Polchinski and Pouliot from a purely perturbative
computation  2) show how various other non-perturbative amplitudes can
be similarly computed, including other M-momentum transferring processes  
and 3) clarify the problem of exhibiting 
the full eleven dimensional Lorentz invariance from the gauge theory 
description.  Something of a similar approach has been discussed in the
context of describing transverse 5-branes in Matrix Theory \cite{GRT,Lif}.  

The recovery of the Polchinski-Pouliot amplitude from a perturbative 
calculation proceeds as follows.  We first note that in ten dimensions the
process corresponds to the exchange of D0-branes between D2-branes.  By
compactifying a transverse direction and performing T-duality, we obtain
 D1-brane exchange between D3-branes.  Finally, S-duality
maps the configuration to the exchange of fundamental strings between
D3-branes, a process which has a purely perturbative gauge theory 
description.  In section 3, we perform the perturbative gauge theory
calculation, which is closely related to Schwinger's calculation of the
pair production of charged particles in an electric field \cite{Sch}, 
implement the 
duality transformations, and show in section 4 that the
result indeed agrees with the  Polchinski-Pouliot amplitude.  

Fundamental string exchange between D3-branes is easily generalized to the
case of other Dp-branes.  Applying duality transformations then allows one
to generate amplitudes for a wide variety of non-perturbative processes,
as we discuss in section 5.  This includes, for instance,
 M-momentum exchange between D0-branes or between D4-branes.  For such
amplitudes, there has been little progress based on the approach of 
directly evaluating the relevant non-perturbative effects.  For instance, the
calculation in the case of D0-branes seems to require detailed knowledge
of the ill-understood bound state wavefunction describing a cluster of 
D0-branes.  However, by using dualities to relate the amplitudes to other
perturbative effects, some progress can be made.  
    
Therefore we see that there is no impediment to computing M-momentum transfer
in a number of different cases.  What does this say about the prospects for
demonstrating eleven dimensional Lorentz invariance in general?  Certainly, a
proof of Lorentz invariance should include supplying a means of computing an
arbitrary M-momentum transferring amplitude.  For our purposes, it
is essential to differentiate between two  types of such
processes: those to which duality can be applied to yield  a perturbative
amplitude, and those which cannot be so transformed.  The results of
\cite{PP} and the present work are evidence that Lorentz invariance is
recovered for the first type of processes, but there is presently no 
evidence regarding the second type.  

  The processes considered here -- and 
every successfully treated example existing in the literature -- are such
that the objects exchanged in a given scattering are of a single type.
This is sufficient to guarantee that the calculations can be reduced to 
perturbation theory with the help of dualities.  This is no longer true
when one consider the exchange of multiple kinds of objects in a given 
amplitude. 
For example, in the case of D3-brane scattering we can compute amplitudes 
corresponding to the exchange of F-strings or of D-strings; however, we cannot
use our techniques to compute the second type of amplitude
which involves the exchange of both F and D strings simultaneously. 
 Specifically, the 
F-string case follows from a perturbative computation, and the D-string case
is then obtained by applying a duality transformation, but an amplitude 
involving the exchange of two strings, one F and one D,
\footnote{Here we do not mean a $(1,1)$ string, which we are able to treat
with our methods, but rather two strings:
a $(1,0)$ and a $(0,1)$.}
 cannot be transformed into a perturbative process.  To demonstrate that the 
latter type of amplitudes are consistent with eleven dimensional Lorentz 
invariance one is forced to do a non-perturbative gauge theory calculation.  
Without explicitly performing such a calculation, it is not possible to 
use dualities to argue for Lorentz invariance in the general case.  As
another example, one can consider the scattering of three membranes with
different amounts of M-momentum transferred between each pair.  Duality
transformations could be used to map a given subprocess to a perturbative
process, but one would still require a non-perturbative calculation to 
correctly evaluate the total amplitude.

The fact that perturbative results plus dualities are enough to establish 
Lorentz invariance for a variety of amplitudes is an encouraging sign that
the symmetry exists in the gauge theory.  In our opinion, though, a true
test of this conjecture will involve evaluating an amplitude which cannot
be made perturbative through any sequence of dualities.  Only then will we
have a result which is not directly implied by perturbation theory plus a
symmetry.

\section{M-momentum transfer in supergravity}

We begin by writing down the type of supergravity amplitude which we wish to 
recover from gauge theory.  For definiteness, we consider the case of 
M-momentum transfer between membranes.  
The idea is to calculate, at the level of classical supergravity, the 
contribution to the potential between two membranes due to the exchange of
momentum in the eleventh direction.  The calculation is a slight modification
of the one appearing in \cite{PP} -- we take the membranes to have a
nonzero relative velocity in the eleventh direction and zero relative 
velocity in the transverse directions, whereas in \cite{PP} the 
opposite configuration is considered.  To perform the calculation we use
the well known source-probe method: we first write down the field 
configuration produced by one of the membranes (the source) and then 
introduce the second membrane by coupling it to the background according
to the supermembrane action.  We then compactify the eleventh direction and
 a transverse direction by replacing the single source membrane by
a periodic array of membranes in the compact directions.
  Finally, we perform a Poisson
resummation to exhibit the sum over exchanged states in a form which is
conveniently compared with gauge theory results.

The metric produced by a membrane in non-compactified eleven dimensional space
\footnote{We take the  coordinates to 
be $t,x^1, \ldots ,x^9, x^{11}$} is \cite{DS}
\be
ds^2 = H_2^{-2/3}[dt^2-(dx^1)^2-(dx^2)^2]-H_2^{1/3}[(dx^3)^2+\cdots + 
(dx^{11})^2].
\ee
\be
H_2=1+\frac{Q_{2}^{(11)}}{r^6} \quad\quad ; \quad\quad
r^2=(x^3)^2+\cdots + (x^{11})^2.
\ee
The action of the probe membrane is given by
\be
S_2=-T_2^{(11)}\int d^3x \left\{ 
\sqrt{\det (g_{MN}\partial_{\mu}X^{M}\partial_{\nu}X^{N})}-H^{(3)} \right\},
\ee
where the determinant acts on the worldvolume indices $\mu,\nu$.
The tension and charge of the membrane are related to the Planck mass by
\be
T_2^{(11)}Q_{2}^{(11)}=\frac{8}{M_{11}^3}
\ee
We take the probe to lie in the $x^0,x^1,x^2$ plane, and to be
 moving with constant velocity $v$ in the eleventh direction.  Evaluating the 
action in static gauge, the leading interaction contribution comes from the 
$v^4$ term
and reads
\be
S_2=\frac{1}{M_{11}^3} \int d^3x \frac{v^4}{r^6}.
\ee
To compactify $x^3,x^{11}$, we replace the single source membrane by a
periodic array:
\be
\frac{1}{r^6}\rightarrow \sum_{n_3,n_{11}} 
\frac{1}{[r^2+(2\pi n_3 R_3)^2 +(2\pi n_{11} R_{11})^2]^3},
\ee
where on the right hand side $r^2=(x^4)^2+\cdots + (x^{9})^2$.
Since we are not presently interested in the exchange of momentum in the
$x^3$ direction, we will take the limit of large $R_3$ and so replace
the sum over $n_3$ by an integral.  In addition, we use the Poisson 
resummation formula to rewrite the sum over $n_{11}$.  We then have,
\be
\frac{1}{r^6}\rightarrow
\frac{3}{16 R_3}\frac{1}{2\pi R_{11}}\sum_{m} \int_{-\infty}^{\infty}
dx \frac{e^{i m x/R_{11}}} {(r^2+x^2)^{5/2}}
= \frac{1}{16 \pi}\frac{1}{R_3 R_{11}^{3}}\frac{1}{r^{2}}
\sum_{m} m^{2} K_{2}(|m|r/R_{11}),
\ee
where $K_{2}$ is a modified Bessel function.
For the action we then find
\be
S_{\rm eff}= \int d^3x 
 \ \frac{1}{16 \pi}\frac{1}{M_{11}^{3}R_3 R_{11}^{3}}
\frac{v^4}{r^2} \sum_{m} m^2 K_{2}(|m|r/R_{11}).
\ee
The $m$'th term in the sum corresponds to the contribution from the 
exchange of $m$ units of M-momentum,  This is seen from the asymptotic
form for large argument of the modified Bessel function
\be
   K_2 ( |m|r/R_{11} ) \sim \sqrt{\frac{\pi R_{11}}{2|m|r}} 
 \ e^{-|m|r/R_{11}} \ ,
\ee
so that the rate of fall off matches with the propagator of a massive 
Kaluza-Klein state.  
Keeping only the leading $m=1$ term, the result for large transverse 
separation is
\be
S_{\rm eff} = \int d^3x 
 \ \frac{1}{16 \sqrt{2\pi}}\frac{1}{M_{11}^{3}R_3 R_{11}^{5/2}}
\frac{v^4}{r^{5/2}} e^{-r/R_{11}}
\label{effaction}
\ee

For comparison with gauge theory, it is helpful to rexpress the result in 
ten dimensional string units using
\be
M_{11}^{3}=\frac{1}{(\alpha ')^{3/2} g_{st}} \quad\quad ;  \quad\quad
R_{11} = \sqrt{\alpha '} g_{st}.
\label{convert}
\ee
In these units,
\be
S_{\rm eff} = \int d^3x 
 \ \frac{1}{16 \sqrt{2\pi}}\frac{(\alpha ')^{1/4}}{g_{st}^{3/2}R_3}
\frac{v^4}{r^{5/2}} e^{-\frac{r}{\sqrt{\alpha '}g_{st}}}
\ee
The appearance of $1/g_{st}$ in the exponential indicates a nonperturbative
effect in the gauge theory description.   In the case of nonzero transverse
velocities, the above formula was reproduced from an instanton calculation
in \cite{PP}.  Our goal in the remainder of this work is to demonstrate
how this and other amplitudes can be obtained starting from purely
perturbative calculations, which are easier to perform.

\section{Gauge theory calculation}

In this section we calculate the contribution to the one-loop gauge theory 
effective action from the exchange of F-strings between bound states of 
D$p$-branes and F-strings.  The calculation is perturbative in gauge theory
and in the string coupling $g_{st}$, though the result will be seen to be
non-perturbative in terms of $\alpha '$.  
In the next section we will show that applying
dualities to the result yields the supergravity amplitude (\ref{effaction}).

The basic description of this process is as follows. Consider two 
D$p$-branes, separated by a transverse distance $r$, and with $q_{1},q_{2}$ 
F-strings bound to them. 
Let one of the $p$ spatial directions  be compact with  length $2\pi R$,
and let the F-strings wind around this direction.  The F-strings are 
described in the gauge theory as electric flux in the unbroken U(1), with
the electric field being proportional to the difference in F-string number, 
$q_1-q_2$.  Massive excitations with U(1) charge in the gauge theory 
correspond to open strings stretching between the two D-branes.  Thus there
is an analogue of the Schwinger effect given by pair creation of open 
strings\footnote{A pair creation effect can also occur on a single
D-brane, if the electric field on it exceeds a critical 
value \cite{BP}.}.  
The  creation of real strings 
gives an imaginary contribution to the effective action \cite{DKPS}. 
If the pair  propagates only in imaginary time
as virtual strings, the contribution to the effective action is real.

Since one of the spatial directions is periodic, the strings (which propagate
in opposite directions) can propagate around the circle $m$ times before
annihilating. Consider a process where they wind once around the circle:

\def\fpsangle{270}
\begin{center}
\leavevmode
\fpsxsize 2in
\fpsysize 2in
\fpsbox{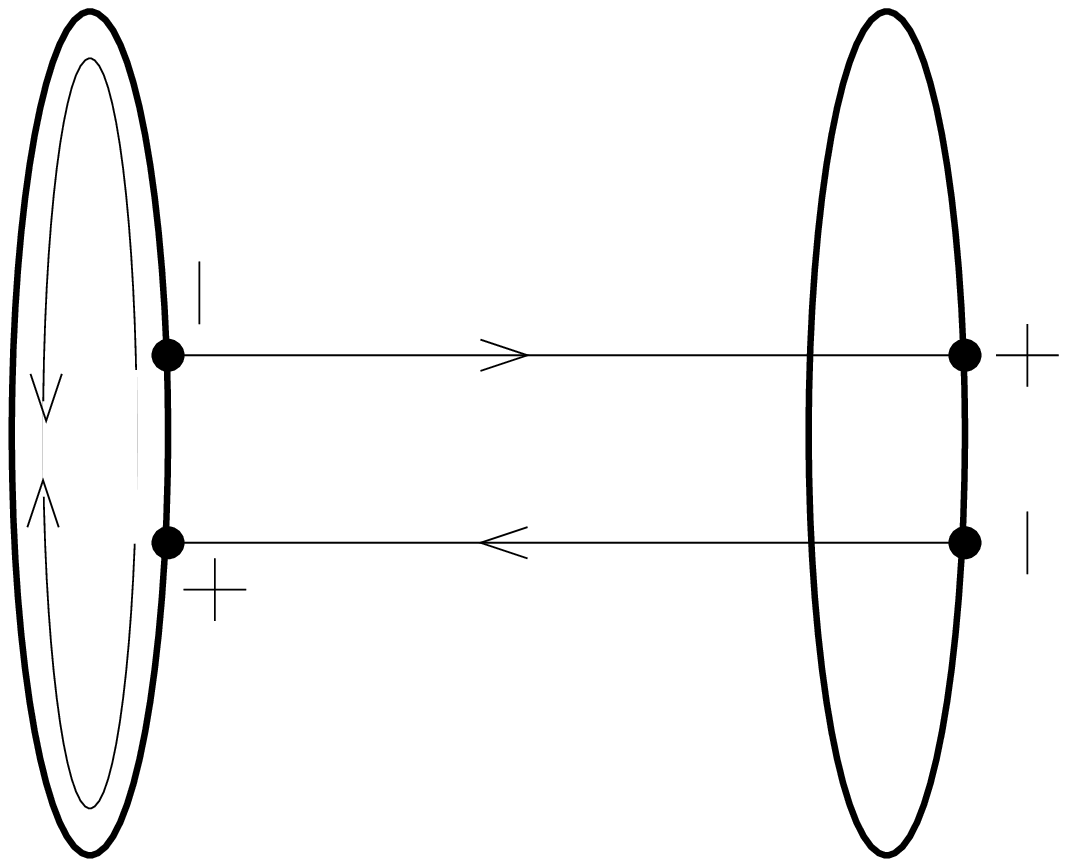}
\end{center}
\vskip 0.1 true cm
\begin{quotation}
\small
\noindent
{\bf Figure :}
F-string transmission. A virtual string pair is created between the
D-branes, propagates around the compact direction and then disappears.
The electric flux lines (not shown) running from positive charges to 
negative charges  will 
remain after the pair has disappeared. Thus the electric flux increases
(decreases) on the upper (lower) brane by one unit,  corresponding to 
F-string transfer from the lower brane to the upper brane.
\end{quotation}
The electric field lines between the endpoints of the string close
around the circle, so the electric flux changes by +1 unit on one D-brane
and by -1 unit on the other D-brane. 
But since the electric flux counts the number of F-strings
bound to the brane, this means that one F-string has transferred from one
D-brane to the other in this process. Similarly, a path
with winding number $m$ transfers $m$ F-strings from D-brane
to D-brane. 

We now move to the gauge theory calculation.
For the most part, the calculation follows the well known method
introduced in \cite{DKPS}. However, there will be a
departure from the standard treatment halfway through  
when we take into account the F-string transmission.

  We consider $p_1 +p_2$ D$p$-branes in the $x^1,\ldots ,x^p$ plane.
We then compactify $x^1,\ldots x^p$ on a torus of radii $R_1,\ldots R_p$, 
and add 
$q_1 + q_2$ F-strings winding along the direction $x^p$. 
The world volume gauge theory is the $D=p+1$
dimensional U($p_1+p_2$) Super-Yang-Mills-Higgs theory with the 
Lagrangian
\be
  \Lcal = \frac{1}{g^2_{YM}} \ \Tr \{ -\frac{1}{4} F^{\mu\nu} F_{\mu
\nu} + \half D^{\mu} \phi_i D_{\mu} \phi_i + \frac{1}{4} [\phi_i,\phi_j]^2 
+ {\rm fermions} \} \ .
\label{Lsym}
\ee

We separate the D$p$-branes by a transverse distance $r$ in 
the $x^D$-direction into two groups with $p_1$ and $p_2$ branes by
 giving the Higgs field $\phi_D$ a vacuum expectation value
\be
 \bar{\phi}_D = \left( \begin{array}{cc} \frac{r}{2} \unitm_{p_1} & \mbox{} \\
                              \mbox{} & -\frac{r}{2} \unitm_{p_2} \end{array}
\right) \ .
\label{bg1}
\ee
This breaks the U($p_1+p_2$) symmetry to U($p_1$)$\times$U($p_2$). 

The $q_1,q_2$ F-strings bound to the
$p_1,p_2$ D-branes are represented by two background electric fields 
\be
    E_1 = q_1 / R_p \ ; \ E_2 = q_2 / R_p \ \ .
\label{Es}
\ee
That is to say, there is a background
gauge field
\begin{eqnarray}
 \bar{A}_{\mu} &=& \zerom_{p_1+p_2}, \ \mu =0\ldots p-1 \ ; \nonumber \\
\bar{A}_p &=& \left( \begin{array}{cc} E_1 t \unitm_{p_1}  & 
\mbox{} \\ \mbox{} & E_2 t \unitm_{p_2} \end{array} \right) \ .
\label{bg2}
\end{eqnarray}
Next, we find the
effective potential between the D-brane bound states by integrating out the
massive field fluctuations about the background
(\ref{bg1}),(\ref{bg2}). We are
interested only in contributions at the one loop level. Since refs. 
\cite{CT,KT} give the result for general D0-brane backgrounds, it is
convenenient to T-dualize our configuration to that of D0-branes, in which
case the gauge field
components $A_i$ are converted to Higgs fields $\phi_i$ by the 
relation \cite{WT,GRT}
$$
   i\pat_i + A_i \leftrightarrow \phi_i \ ,
$$ 
The resulting effective action $S_{\rm eff}$ takes the following
functional determinant form:
\begin{eqnarray}
 e^{iS_{\rm eff}} &=& \det (H-i2E)^{-1} \det (H+i2E)^{-1} 
   \det (H)^{-6} \nonumber \\
        \mbox{} &\mbox{}&  \det (H+iE)^4 \det (H-iE)^4 \label{W} 
\end{eqnarray}
where the operator $H$ is
\be
 H = [-\pat^2_t  +\pat^2_1 +\cdots 
       +\pat^2_{p-1} + (\pat_p - iEt)^2 + r^2]\cdot \unitm_{p_1p_2} 
\label{H}
\ee
and $E$ is the difference of the two electric fields $E_1,E_2$,
\be
         E = [E_1 - E_2]\cdot \unitm_{p_1p_2} \ .
\ee
As expected, only the difference of the electric fields appears.
Note that in contrast to the usual practise in the literature,
in the above we work in real time $t$. We now
use Schwinger's trick to exponentiate the determinants:
\be
  S_{\rm eff} = ip_1p_2 \int^{\infty}_0 \frac{ds}{s} \ (6+2\cosh 2Es 
  - 8 \cosh Es ) \ \tr e^{-iHs} \ .
\label{Wsch}
\ee
(Note that now $H,E$ no longer contain 
the unit matrix $\unitm_{p_1p_2}$ -- it has been traced over
to give the factor $p_1p_2$.)

To this point the calculation has been standard procedure;  the new twist
comes from evaluating $tr e^{-iHs}$.
In the coordinate representation, the trace becomes an integral over
a quantum mechanical propagator $K(x,y;s)$:
\be
 \tr e^{-iHs} = \int d^D x \ \bra x | e^{-iHs} | x \ket
              \equiv \int d^D x \ K(x,x;s) \ .
\label{trace}
\ee
The operator $H$ is identical to the small fluctuation operator for a
massive charged scalar field minimally coupled to a uniform background
electric field $E$. 

In noncompact space, the propagator  $K_H(x,y;s)$ for
a charged particle in an electric field
can be found in  {\em e.g.} \cite{Stephens,BD},
\be
 K_H(x,y;s) = \frac{iE}{(4\pi i)^{D/2} s^{D/2 -1} 
  \sinh (Es)} \ e^{-ir^2s +i(E/4)[\coth (Es)
((y^p-x^p)^2 -(y^0-x^0)^2) + 2(x^p y^0-x^0y^p)]} \ .
\label{kernel}
\ee
For a path corresponding to an instanton we take 
\be
      x^0=y^0=t.
\label{time} 
\ee
Since we take space to be
 a torus, the propagator $K(x,x;s)$ in (\ref{trace}) is a sum of
 contributions with different winding number $m \in {\bf Z}$
corresponding to paths which wind $m$ times around the $x_p$-cycle:
\be
     K(x,x;s) = \sum^{\infty}_{m=-\infty} K_H(x,x+2\pi m R_p e_p ;s)  
\label{windings}
\ee
where $K_H$ is the propagator (\ref{kernel}) with (\ref{time}):
\be
    K_H(x,x+2\pi m R_p;s) = \frac{iE}{(4\pi i)^{D/2} s^{D/2 -1} \sinh (Es)} 
    \ e^{-ir^2 s + i E m^2 \pi^2 R^2 \coth (Es)-iE\pi mRt} \ .
\ee
Then, after rescaling $s\rightarrow s/r^2$, 
the effective action reads as follows, 
\begin{eqnarray}
   S_{\rm eff}  &=& ip_1p_2\int d^D x \int^{\infty}_0 ds
 \ \frac{ iE(6 + 2\cosh (2Es/r^2) -
 8 \cosh (Es/r^2) ) }{(4\pi i s)^{D/2} 
   \sinh (Es/r^2)} \nonumber \\
         \mbox{} &\mbox{} & \ \ \ \ \ \ \ \ \sum^{\infty}_{m=-\infty} 
      \exp \{ -is + iE m^2 \pi^2 R^2_p \coth (Es/r^2) -iE\pi mR_pt \} \ .
\label{W1}
\end{eqnarray}
Assuming that the bound states are well separated, 
$$
  \frac{E}{r^2} \ll 1 \ , 
$$
so we can expand in $E/r^2$. we find the leading contribution
to the effective action 
\be
   S_{\rm eff}  = \int d^D x \ \frac{\pi^{4-D} p_1p_2 E^4 R_p^{4-(D/2)}}
  {2^{D-1} r^{4-(D/2)}}  \sum^{\infty}_{m=-\infty}
 |m|^{4-(D/2)}  \ K_{4-(D/2)}(2\pi |m| R_p r) \ e^{-iE\pi mR_p t} \ \ ,
\label{Veff}
\ee
where $K_{4-(D/2)}$ is a modified Bessel function, with asymptotic behavior
for large argument\footnote{One can also check that shrinking a compact
transverse direction to zero changes the power law falloff of (\ref{Veff})
correctly for the lower dimension. For zero-brane scattering,
this was discussed in \cite{BC}.}
$$
  K_{4-(D/2)} (2\pi |m| R_p r) \sim \frac{1}{2\sqrt{|m| R_p r }}
                      \ e^{-2\pi |m| R_p r} \ .
$$

In the above, we have set $2\pi \alpha' = 1$. In the next section,
it will be useful to have  the result for D3-branes $(D=4)$, keeping track of
correct dimensions. After restoring $\alpha'$, and 
setting $p_i=m=1,\ t=0$ for simplicity, the asymptotic form of the effective
action becomes
\be
 S_{\rm eff} = \int \frac{d^4 x}{(2\pi \alpha')^{2}}
 \ \frac{ (2\pi \alpha' E)^4 R_3^{3/2} \sqrt{2\pi \alpha'}}{16 r^{5/2}}
  \ e^{-R_3 r/\alpha' } \ .
\label{Veff3}
\ee
The appearance of $1/\alpha'$ in the exponent indicates that the result
is non-perturbative in terms of $\alpha'$.

\section{M-momentum transfer as F-string transfer}

We now show how to recover the non-perturbative result for M-momentum
exchange between membranes starting from the perturbative process of
fundamental string exchange between D3-branes.  We begin with two  D3-branes
in the $x^1,x^2,x^3$ plane, where $x^3$ is compactified on a circle of radius 
$R_3$. The D3-branes have worldvolume electric fields pointing in the
$x^3$ direction; we denote $E=E_1-E_2$ as the difference in the two electric
fields.  The effective potential due to the exchange of F-strings wrapped
around the $x^3$ direction is given by (\ref{Veff3}).  We will perform the
following  duality transformations:

\mbox{}

\noindent
1)   S-duality.  Process then corresponds to exchange of D-strings wrapped on
on $x^3$ between D3-branes with relative magnetic fields $B=E$.

\mbox{}

\noindent
2) T-duality in $x^3$ direction.   Process then corresponds to exchange of
D0-branes between D2-branes with relative magnetic fields $B$.

\mbox{}

The final process is precisely M-momentum transfer between membranes, and is
a non-perturbative effect arising from instantons.  

The first step, S-duality, takes $g_{st} \rightarrow 1/g_{st}$, 
$E  \rightarrow B$, and rescales lengths as measured in the string metric:
$x \rightarrow x/\sqrt{g_{st}}$.  We will express results in terms of the
radius of the eleventh dimension, $R_{11}$, using (\ref{convert}).
We obtain for the action
\be
S_{\rm eff}=\int d^4x \frac{(\alpha')^{3/4}}{16(2\pi \alpha ')^{3/2}}
\frac{R_3^{3/2}}{R_{11}^{3/2}} \frac{(2\pi \alpha ' B)^4}{r^{5/2}}
e^{-\frac{R_3 r}{\sqrt{\alpha '}R_{11}}}
\ee
The second step, T-duality in direction $x^3$, takes
\be
R_3 \rightarrow \frac{\alpha '}{R_3} \quad\quad ; \quad\quad 
R_{11} \rightarrow \frac{\sqrt{\alpha '}}{R_3}R_{11}.
\ee
The action now becomes
\be
S_{\rm eff}=\int d^3x \frac{(\alpha ')^{3/2}}{16\sqrt{2\pi \alpha '}}
\frac{1}{R_{11}^{3/2}R_3}\frac{(2\pi \alpha ' B)^4}{r^{5/2}}
e^{-\frac{r}{R_{11}}}
\ee
where the integral over the $x^3$ direction has been performed to give a factor
of $2\pi (\alpha '/R_3)$.

Finally, we rexpress the result in terms of the eleven dimensional Planck
mass using (\ref{convert}),
and reinterpret the magnetic field as velocity in the eleventh direction
\be
2\pi \alpha' B \rightarrow v.
\ee
Our final result for the effective action is then
\be
S_{\rm eff}= \int d^3x \frac{1}{16\sqrt{2\pi}} \frac{1}{M_{11}^3 R_{3}
R_{11}^{5/2}
} \frac{v^4}{r^{5/2}} e^{-\frac{r}{R_{11}}}.
\label{Vfinal}
\ee
we see that the result agrees with (\ref{effaction}).

\section{Other transfer processes}

After having shown that nonperturbative M-momentum transfer between
D2-branes
can be related to the  perturbative F-string transmission process,
a natural question to ask is what other processes can be understood
by relating them to the F-string transmission  by various dualities.
In this section we discuss some other examples.

Let us take as a starting point the case $p=1$ of section 2: transfer
of m F-strings between $(p,q)$-strings in IIB theory. 
Recall that $(p,q)$-strings are bound states of $p$ D-strings and $q$
F-strings and that they are stable against decay into multiple strings
if $p$ and $q$ are relatively prime: $\gcd (p,q)=1$ \cite{pq}.
Then, note that any such
$(p,q)$-string can always be mapped to a fundamental string by
an SL(2,Z) transformation:
\be
     \left( \begin{array}{cc} q & -p \\ a & b \end{array} \right)
     \left( \begin{array}{c} p \\ q \end{array} \right) 
   = \left( \begin{array}{c} 0 \\ 1 \end{array} \right) \ .
\label{map}
\ee
The integers $a,b$ such that 
\be
     ap + bq = 1 
\ee
can be found using Euclid's algorithm. Obviously, $m$ (p,q)-strings
map to $m$ F-strings. Therefore, a $(p,q)$-string
transmission process
\be
        (p_1,q_1) \stackrel{(p,q)}{\longleftarrow \longrightarrow} (p_2,q_2)
\label{proc1}
\ee
(meaning $(p,q)$-strings transferred between a 
$(p_1,q_1)$-string and a $(p_2,q_2)$-string) is related to the simple
F-string transmission process\footnote{Some restrictions apply here: we
assume that $qp_i-pq_i>0$ so that after the SL(2,Z) transformation 
(\ref{map}) the images of the bound states $(p_i,q_i)$ have a strictly
positive D-string charge. Then we can avoid dealing with pairs of D-strings
and D-antistrings.} of 
section 2 by a transformation (\ref{map}).
\footnote{A interesting case to consider is when the final string has 
$(p_2,q_2)$
{\em not} relatively prime.  In this case  one could end up with 
either with a marginally bound $(p_2,q_2)$ string, or a multiple string state
corresponding to its constituents.  How to distinguish such final states in
our formalism is not clear.}

Suppose that the $(p,q)$-strings wind around compact direction $x_1$. T-duality
along this direction then maps the process (\ref{proc1}) to one in which
 D0-branes are transferred between clusters of D0-branes: 
\be 
    p_1 D0 + v_1  \stackrel{p D0 + v}{\longleftarrow \longrightarrow}
    p_2 D0 + v_2 \ ,
\label{proc1b}
\ee
with transverse velocities $v_i=q_i/R_1$ and $v=q/R_1$ where $R_1$ is 
the radius of the compact direction $x_1$. In other 
words, we can calculate M-momentum transfer between gravitons.  Such a
calculation seems to be very difficult to perform directly without 
a detailed understanding of the properties of marginally bound states of
D0-branes.  

Now suppose 
that the directions $x_2,x_3$ are also compact. Then, T-duality in directions
$x_2,x_3$ ($T_2T_3$ for short) relates the process (\ref{proc1}) to
\be
        p_1D3+q_1F1 \stackrel{pD3+qF1}{\longleftarrow \leftharpoondown 
                 \rightharpoondown \longrightarrow} p_2D3+q_2F1 \ .
\label{proc2}
\ee
 (\ref{proc2}) is the generalized version of the F-string transfer of
section 2: the bound states can exchange D3-branes as well as F-strings.

Continuing with S-duality and T-duality $T_1$, (\ref{proc2}) in turn
maps to a more general version of the M-momentum transfer process of \cite{PP}:
\be
     p_1D2+q_1D0 \stackrel{pD2+qD0}{\mapb} p_2D2+q_2D0 \ ,
\label{proc3}
\ee
where both D2-branes and D0-branes can be transferred.

Another T-duality $T_2$ maps (\ref{proc3}) to a process involving diagonal
D-strings winding $p,q$ times along the directions $x_3,x_2$. We use the
homology notation $p(3)+q(2)$ for such diagonal strings \cite{WTlect}. 
We obtain
\be
 p_1(3)+q_1(2)\ D1 \stackrel{p(3)+q(2)\ D1}{\mapb} p_2(3)+q_2(2) \ D1 
\label{proc4}
\ee
where the diagonal D1-strings change their winding.

Now, if we go back to our starting point, the F-string transmission, we
realize that we could just as easily  have taken the strings to be diagonal
in target space with winding $m(1)+n(2)$ along the $x_1,x_2$-cycles. 
It would be very easy to repeat the calculation of section 2 in this case.
The SL(2,Z) transformation (\ref{map}) relates this to a diagonal
version of (\ref{proc1}).
The power of introducing diagonal winding is that by compactifying $x_4$
and by using a sequence of dualities $T_3T_4ST_1$, we can end up with a
quite complicated process 
\begin{eqnarray}
  p_1m(1234)D4+q_1m(12)D2+p_1n(34)D2+q_1nD0 \nonumber \\
 \stackrel{pm(1234)D4+qm(12)D2+pn(34)D2+qnD0}{\mapc}
  p_2m(1234)D4+q_2m(12)D2+p_2n(34)D2+q_2nD0
\label{proc6}
\end{eqnarray}
involving D(4+2+0)-transfer between D(4+2+0)-branes. Note that in particular
it involves
M-momentum transfer between D4-branes\footnote{In this
particular case M-momentum transfer always comes along with some D2-brane
transfer due to the manner in which lower dimensional branes
bind to D4-branes.}. It is not obvious how to accomplish this
by approaching the problem as one of finding a suitable non-perturbative 
configuration in 4+1-dimensional field theory. 

So, we conclude that a variety of non-perturbative transfer processes can be
analyzed quantitatively by exploiting  dualities to relate them to  the simpler
perturbative process of electric flux
transfer between D-branes.  In particular, one can study M-momentum
transfer between gravitons (\ref{proc1b}), membranes (\ref{proc3}) and
longitudinal fivebranes (\ref{proc6}) via this route.

\bigskip

{\Large {\bf Acknowledgements}}

\bigskip

We would like to thank M.~Aganagic, S.~Cherkis, E.~Gimon, and T.~K\"{a}rki
for discussions. We are grateful for the hospitality of the 
Institute for Theoretical Physics in Santa Barbara, where a part of
this work was carried out. This research was supported in part by the
National Science Foundation under Grant No. PHY94-07194.

\end{document}